\documentclass[acmsmall]{acmart}
\usepackage{multirow}
\usepackage{cleveref}
\usepackage{csquotes}
\usepackage{xcolor}
\usepackage{booktabs}
\usepackage{subfig}
\usepackage[export]{adjustbox}
\AtBeginDocument{%
  }

\setcopyright{cc}
\setcctype{by}
\acmJournal{PACMHCI}
\acmYear{2025}
\acmVolume{9}
\acmNumber{7}
\acmArticle{CSCW490}
\acmMonth{11}
\acmDOI{10.1145/3757671}

\begin{document}

\title{The Future of Tech Labor: How Workers are Organizing and Transforming the Computing Industry}

\author{Cella M. Sum}
\email{csum@andrew.cmu.edu}
\affiliation{%
  \institution{Carnegie Mellon University}
  \city{Pittsburgh}
  \state{Pennsylvania}
  \country{USA}
}

\author{Anna Konvicka}
\email{ak6206@princeton.edu }
\affiliation{%
  \institution{Princeton University}
  \city{Princeton}
  \state{New Jersey}
  \country{USA}
}

\author{Mona Wang}
\email{monaw@princeton.edu}
\affiliation{%
  \institution{Princeton University}
  \city{Princeton}
  \state{New Jersey}
  \country{USA}
}

\author{Sarah E. Fox}
\email{sarahf@andrew.cmu.edu}
\affiliation{%
  \institution{Carnegie Mellon University}
  \city{Pittsburgh}
  \state{Pennsylvania}
  \country{USA}
}

\renewcommand{\shortauthors}{Cella M. Sum, Anna Konvicka, Mona Wang, and Sarah E. Fox}

\begin{abstract}
The tech industry's shifting landscape and the growing precarity of its labor force have spurred unionization efforts among tech workers. These workers turn to collective action to improve their working conditions and to protest unethical practices within their workplaces. To better understand this movement, we interviewed 44 U.S.-based tech worker-organizers to examine their motivations, strategies, challenges, and future visions for labor organizing. These workers included engineers, product managers, customer support specialists, QA analysts, logistics workers, gig workers, and union staff organizers. Our findings reveal that, contrary to popular narratives of prestige and privilege within the tech industry, tech workers face fragmented and unstable work environments which contribute to their disempowerment and hinder their organizing efforts. Despite these difficulties, organizers are laying the groundwork for a more resilient tech worker movement through community building and expanding political consciousness. By situating these dynamics within broader structural and ideological forces, we identify ways for the CSCW community to build solidarity with tech workers who are materially transforming our field through their organizing efforts.
\end{abstract}



\received{October 2024}
\received[revised]{April 2025}
\received[accepted]{August 2025}

\maketitle

\section{Introduction}
The tech industry has long cultivated a mythos of meritocracy and innovation---an elite world of generous salaries, creative autonomy, and world-changing impact. This image has shaped public perception, drawn in workers, and helped legitimize tech’s expanding influence over society. Yet, beneath this carefully crafted narrative lies a much harsher reality. Workers across the tech supply chain---from full-time engineers and product managers to contractors, logistics staff, and platform-based gig workers---now face mass layoffs, algorithmic surveillance, opaque management structures, and diminished job security \cite{so2025cruel, sannon_2022,dubal2023algorithmic}.
In response, U.S.-based tech workers are organizing---from union drives at Kickstarter and among Google subcontractors, to grassroots campaigns like No Tech for Apartheid ~\cite{2024_tan_unlikely}. These efforts challenge not only exploitative conditions, but also deeper structural and cultural forces that shape how tech work is configured and understood. Wages and job security are not the only things at stake. Tech workers are fighting for a radically different vision of how the tech industry functions and who it serves.

Recent CSCW and HCI research has begun to trace these tensions by documenting the precarity and stratification of computing labor as it is increasingly shaped by outsourcing, algorithmic management, and venture capital restructuring \cite{gray2019ghost, kellogg2020algorithms, 2024_shestakofsky_startup}. Other work examines tech workers' resistance to these dynamics through grassroots organizing, mutual aid, and digital tools that support coalition building and advocacy \cite{qadri2021s, thuppilikkat2024union, lee2015working, bonini2024cooperative, yao2021together}. Examples of this work include platforms like Turkopticon and Dynamo that facilitate collective visibility and negotiation \cite{irani_2013, harmon2019rating, salehi2015salehi}, as well as community-driven repositories like \textit{Collective Action in Tech} that track the growing movement of tech worker resistance \cite{2024_tan_unlikely}. While prior work highlights the forces shaping tech labor and the tools for resistance, less is known about the lived experiences and subjectivities of workers who have organized across the tech supply chain.

We build on this literature by centering the voices of 44 U.S.-based tech worker-organizers. Through in-depth interviews, we examine how these workers navigate disempowerment in the workplace and how they develop strategies for collective action. Our study is guided by four key research questions: 
\textit{(1) What are the motivations that drive these workers to unionize? (2) What challenges do these individuals face in organizing, and what are their strategies to circumvent them? (3) How do these experiences shape their identities as tech workers and perceptions of the tech industry? (4) What are their visions for building a stronger tech worker movement?} 

Mirroring how organizing spaces strategically define “tech worker” as an umbrella term to foster solidarity \cite{niebler2025tech, 2025_circuit_breakers}, we adopt a broad definition that includes engineers, product managers, customer support specialists, QA analysts, logistics staff, platform-based gig workers, as well as staff union organizers involved in tech union campaigns. Our findings reveal a complex picture of workers managing to build solidarities across roles and geographies despite facing increasingly fissured and unstable workplaces. Through these efforts, they also challenge dominant narratives by presenting tech workers as collective agents of change rather than autonomous individuals.

To interpret these dynamics, we draw on the \textit{Californian Ideology} \cite{barbrook1996californian}---a techno-libertarian worldview that celebrates individual genius, entrepreneurial risk-taking, and market-based solutions, while obscuring the structural inequities that underpin the tech industry. While not the root cause of labor precarity, this ideology functions as a powerful cultural force that discourages organizing by positioning workers as self-reliant innovators rather than participants in a shared struggle \cite{tarnoff2020techworker}. We situate it within a broader political economy shaped by platformization, weak labor protections, and financialization \cite{2024_shestakofsky_startup, 2019_rahman_thelen_platform_business}, all of which constrain the possibilities for collective agency.

Our contribution to the CSCW community is twofold.  First, we extend literature on computing labor and collective action by foregrounding the perspectives of U.S.-based rank-and-file  organizers across the tech supply chain. Second, we argue that the CSCW and HCI communities---often positioned as adjacent to but separate from the tech labor movement---must recognize their entanglement in the industry's material conditions. In doing so, we echo calls for a more politically engaged and worker-centered research agenda \cite{2020_fox_worker, 1996_greenbaum_labor, 2023_tang_cscw_labor}, one that supports organizing not only through tool design, but also through active solidarity and institutional change. While situated in the U.S., our study surfaces dynamics---precarity, stratification, and fissuring---that are increasingly transnational in scope. We hope this work contributes to broader conversations about labor in computing and the urgent need to build cross-border solidarities for more just technological futures.  

\section{Related Work}
In the following section, we outline the related work in two areas. We first examine the infrastructures and ideologies that contribute to an increasingly fissured and disempowered tech labor force. We then trace the histories and strategies of collective organizing in the industry and examine how HCI research has engaged with these efforts.

\subsection{Infrastructures and Ideologies of the Tech Industry}

The labor force in the tech industry is highly stratified---consisting of higher-paid computing professionals such as engineers, programmers, and data analysts alongside lower-paid contingent workers such as warehouse workers, gig workers, and data annotators \cite{tarnoff2020techworker}. In the U.S., tech companies are increasingly turning to subcontracting and outsourcing \cite{brady2023ghost}, shifting both service and computing roles to third-party vendors. This strategy allows firms to cut costs by employing contractors on a temporary basis with lower wages and fewer benefits compared to their full-time counterparts \cite{tech2021separate}. Meanwhile, the rise of platform-based gig work has further fragmented the labor force, subjecting ride-hail and delivery drivers, freelancers, and data workers to surveillance, algorithmic control, and job insecurity with diminished labor protections \cite{kellogg2020algorithms, dubal2023algorithmic, lee2015working, kinder2019gig, kusk2022working, sannon_2022, lampinen2018power, seetharaman2021delivery, silberman2018responsible}. 

In addition, U.S. tech companies are increasingly relying on offshored labor in regions with lower wages and weaker labor laws, particularly in the Global South. For example, companies such as OpenAI and Google rely on data annotators in countries such as Kenya, India, and the Philippines to label vast amounts of training data for AI systems \cite{perrigo2023openai, wang2022aidreamsearchaspiration, dzieza2023inside}. These workers are themselves at risk of being displaced as the industry pursues developments in automation aimed at reducing the need for human oversight \cite{kapania2025examining, prada2024tiktok, xiang2023chatgpt}. The rapid adoption of AI in the tech workplace has also led to mass layoffs in traditional office-based roles, not only affecting customer support and quality assurance roles, but also engineering \cite{robson2025tech, so2025cruel}. Further, foreign nationals on H-1B visas make up a significant portion of the U.S. tech workforce \cite{ingram2025h1b}. Their immigration status is directly tied to their employer,  which limits their ability to switch jobs or speak out against their working conditions, making them more vulnerable to exploitation \cite{twc2025immigrant}.

These developments reflect a convergence of political-economic forces---including the rise of venture capital-backed and platform-based business models---that have restructured the tech labor force \cite{2019_rahman_thelen_platform_business, 2024_shestakofsky_startup}. Yet, the cultural narratives that dominate the tech industry often obscure these inequalities. One such narrative is the \textit{Californian Ideology} (CI), a techno-libertarian worldview coined by Barbrook and Cameron in the 1990s \cite{barbrook1996californian}. Rooted in Silicon Valley’s early entrepreneurial culture, this ideology merges countercultural individualism with neoliberal market logic, celebrating personal innovation, meritocracy, and technological progress while resisting regulation or structural change. As Barbrook and Cameron argue, this vision \emph{“depends upon a willful blindness towards the other} \cite{barbrook1996californian}.” While their original critique was in reference to the racism and inequities occurring in Silicon Valley at the time, we argue that this ideological formation persists within contemporary tech culture by obscuring the labor that sustains digital infrastructures. This includes outsourced, gig, and contingent workers---often women, migrants, or people in the Global South---whose essential labor is nevertheless rendered invisible \cite{gray2019ghost, irani2015cultural}.

These structural and ideological forces also influence how the tech industry approaches questions of ethics and governance. Metcalf et al. \cite{metcalf2019owning} argue that the industry’s reliance on meritocracy, technological solutionism, and market fundamentalism enables it to sidestep regulation and structural change by framing ethics as a technical problem solvable through internal evaluative frameworks \cite{wong2023seeing}. Widder and colleagues \cite{widder2023dislocated} find that the modular and dislocated nature of the tech supply chain causes engineers to feel alienated from the downstream consequences of their work. Further, CI’s emphasis on individualism and technical problem-solving fosters a cultural resistance to organizing by encouraging workers to see success and innovation as a personal achievement rather than a collective struggle \cite{tarnoff2020techworker}. Our study expands this discourse by foregrounding worker experiences to show how these frameworks manifest in everyday organizing challenges across a stratified and fragmented tech workforce.

\subsection{Collective Organizing Across the Tech Pipeline}

Tech workers in the U.S. and globally have a long, if often under-recognized, history of collective action. While unionization in computing has historically been difficult, workers across the tech pipeline have consistently organized around ethical concerns, poor working conditions, and structural inequities \cite{tarnoff2020techworker}. Tech worker organizing in the U.S. emerged in the 1960s--70s, when groups like Computer People for Peace protested the Vietnam War at ACM conferences ~\cite{2019_kagan_greenbaum}, and workplace-based groups like the IBM Black Workers Alliance and the Polaroid Revolutionary Workers Movement mobilized around racial justice and anti-apartheid campaigns~\cite{2022_haeyong_ibm_black_workers, 2020_polaroid}. 
Internationally, mass strikes in countries like France and Italy led to sectoral collective bargaining in the tech industry \cite{andrews2019lessons}, while Scandinavian approaches to participatory design created models for collaborative worker-driven technology development \cite{bransler1989systems, ehn1988work, iivari1998research}.
Despite these global developments, no U.S.-based tech union succeeded during this period~\cite{1974_garner_computer_workers, 1978_gilbert_organizing, greenbaum2020questioning}. 

In the U.S., the structure of labor law poses significant obstacles. Unions are typically recognized at the company level and must win majority support through elections overseen by the National Labor Relations Board (NLRB)\footnote{The National Labor Relations Board (NLRB) is a U.S. government body that oversees union elections and certifications and that also enforces labor laws, including unfair labor practices.}. This company-specific model contrasts with sectoral bargaining or worker councils found elsewhere \cite{thuppilikkat2024union, harmon2019rating, gerbracht2024codetermination}. As a result, U.S. workers in fissured tech environments face steep barriers to traditional unionization \cite{milkman2020union}. Still, worker organizing has adapted to these constraints. For example, gig workers, who are often classified as independent contractors \cite{bales2016uber, tassinari2020riders}, have developed alternative models such as mutual aid networks \cite{qadri2021s, qadri2021mutual, irani_2013}, worker cooperatives \cite{salvagni2022gendering, grohmann2022beyond, lampinen2018member}, and grassroots coalitions \cite{coworker_gig_workers_collective, rideshare_drivers_united}. Communication platforms like WhatsApp, online forums, and social media have become critical spaces for gig workers to share knowledge, provide peer support, and mobilize \cite{qadri2021s, thuppilikkat2024union, lee2015working, bonini2024cooperative, yao2021together, abilio2021struggles, qadri2021mutual, sannon_2022, seetharaman2021delivery}.

While these alternative forms of organizing illustrate how workers navigate the constraints of U.S. labor law today, they are part of a broader landscape of labor activism in tech. Starting in the 1990s, subcontracted service workers—including shuttle drivers, security guards, and cafeteria staff—began organizing in Silicon Valley with support from unions like SEIU and UNITE HERE \cite{tarnoff2020techworker, weigel2017coders}. Later, a renewed wave of activism emerged during the first Trump administration, spurred by \#MeToo, Black Lives Matter, and growing discontent with tech's complicity in state surveillance, climate inaction, and militarism \cite{tarnoff2020techworker, su2021critical}. High-profile examples include the 2018 Google Walkout \cite{chiu2018google, wong2019google}, petitions against Project Maven, and the No Tech for Apartheid campaign ~\cite{anonymous_2021_nimbus, 2022_harrington_nota, shane2018google}. In 2019, subcontracted Google workers at HCL America won the first NLRB-recognized tech union in the U.S.~\cite{2019_ghaffary_hcl}. This was soon followed by the successful Kickstarter Union campaign ~\cite{2023_stolzoff_kickstarter} and the formation of the Alphabet Workers Union as a minority union\footnote{A minority union represents a non-majority of workers within a workplace who voluntarily join and pay dues but cannot negotiate collective bargaining agreements.}~\cite{isaac2021google}. More recently, Amazon warehouse workers at the JFK8 facility in New York formed the company's first NLRB-recognized union ~\cite{new_york_times_2022_amazon}, while video game workers launched United Videogame Workers, an industry-wide video game union~\cite{2025_cwa-unitedvideo}. National unions such as Communication Workers of America (CWA), the Office and Professional Employees International Union (OPEIU), and the Teamsters have all launched coordinated efforts to organize the tech sector~\cite{2020_cwa-code_launch, 2021_opieu_1010_launch,amazon2025teamsters}. 

Retaliation (including firings and mass layoffs) ~\cite{2024_thornbecke_google, 2024_techcrunch_layoffs} has become increasingly common, signaling the intensifying risks that organizers face \cite{bbc2018google}. Despite these increasing risks, many tech workers remain motivated by a growing awareness of social and ethical issues. Recent research suggests that participation in social and ethical activism \cite{2024_nedzhvetskaya_ai_power,2022_nedzhvetskaya_ai_governance, 2022_boag_tech_organizing, widder2023pwer, widder2024power} strongly predicts tech workers' engagement in labor organizing ~\cite{2024_tan_unlikely}, and that organizing can be both overt and subtle, including tactics that leverage corporate discourse and demand change from within \cite{2021_wong_tactics}. Although global in scope, most tech organizing still occurs at the local or company level \cite{niebler2023transcending}. However, some notable forms of cross-border solidarity have emerged. Niebler finds that transnational tech organizing---such as Tech Workers Coalition, the Google Walkouts, and U.S.-based support of the 996.ICU movement in China---operates primarily through informal, grassroots, and digitally mediated networks, outside traditional union frameworks \cite{niebler2023transcending}. These efforts underscore the importance of digital infrastructure and political community in sustaining distributed movements. 

Calls to action within HCI and CSCW communities emerged as early as 1996, when Greenbaum argued that CSCW's focus on work should be expanded to labor issues~\cite{1996_greenbaum_labor}. In recent years, HCI researchers have increasingly focused on designing tools for labor organizing \cite{irani_2013, harmon2019rating, salehi2015salehi}. Platforms like Turkopticon and Dynamo enable Amazon Mechanical Turk workers to coordinate, share employer reviews, and collectively resist poor conditions \cite{irani_2013, salehi2015salehi}. Scholars have also called for worker-centered design approaches \cite{2020_fox_worker}, as well as broader forms of “digital workerism” that support worker agency through data and collective infrastructure \cite{calacci_2022, calacci_2023}. Our study builds on this work by foregrounding the experiences of tech workers across various roles and organizational contexts, revealing emergent strategies and persistent obstacles in building durable labor power in an increasingly fragmented industry. 

\section{Methodology}

\subsection{Interviews}
\subsubsection{Recruitment}
\label{sec:recruitment}
We conducted semi-structured interviews with 44 U.S.-based tech worker-organizers across 28 different workplaces. We define “tech worker” as including workers in technical roles at non-technical companies (e.g. engineers at media or non-profit companies) and workers in non-technical roles at tech companies (e.g. customer support or logistics workers). This includes contractors and platform-based gig workers, in addition to full-time employees. We define “organizer” as any rank-and-file worker organizing collective actions or NLRB-recognized unions, as well as staff union organizers involved in tech organizing campaigns. We recruited participants through convenience and snowball sampling, over social media, on the Tech Workers Coalition Slack channel with permission, and by distributing flyers at a tech worker organizing conference. Interviewees were paid \$20 for their participation. The interviews, which lasted about 90 minutes, were audio-recorded with consent, auto-transcribed, and manually reviewed.

\subsubsection{Participant Composition and Limitations}

\begin{figure}%
\subfloat[Type of Workplace]{
\includegraphics[width=0.6\textwidth,valign=c]{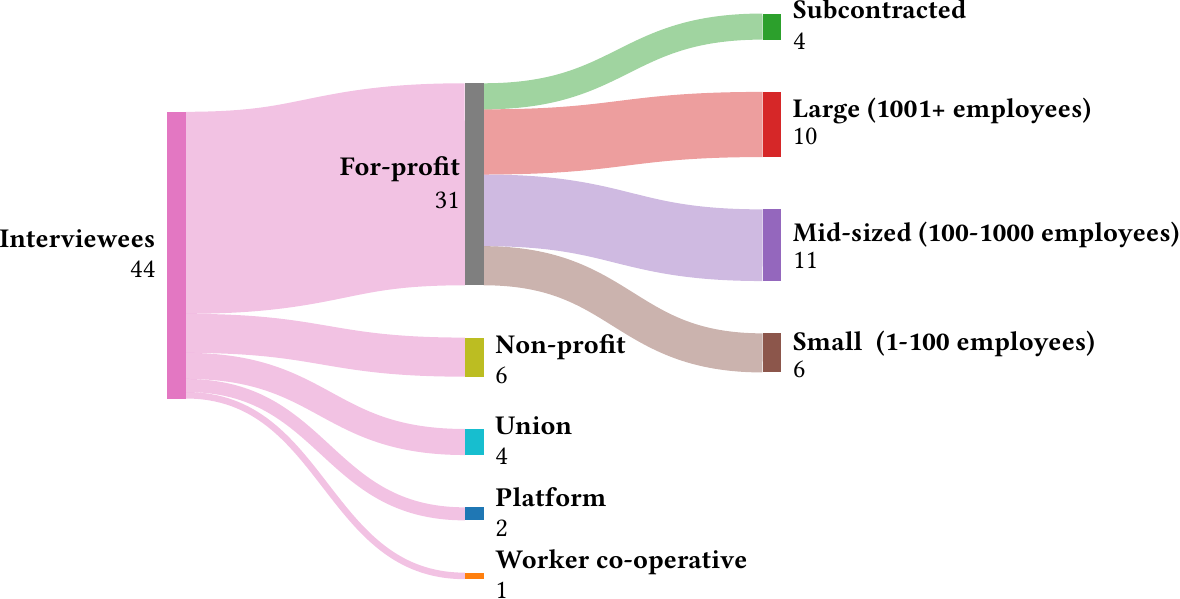}
\Description[Sankey diagram breaking down the workplace composition of our 44 interview pariticpants.]{Sankey diagram breaking down the workplace composition of our 44 interview pariticpants. 31 worked at for-profits, 6 at non-profits, 4 at unions, 2 at platform companies, and 1 at a worker co-operative. Of the 31 for-profit workers, 4 worked at companies subcontracted by a larger tech company, 10 worked at large companies with 1001+ employees, 11 worked at mid-sized companies with 101-1000 employees, and 6 worked at small companies with under 100 employees.}
}
  \quad
\subfloat[Type of role]{
    \adjustbox{valign=c}{
        \footnotesize
        \begin{tabular}{rl}
            \toprule
            Engineer & 22 \\ \hline
            Non-engineer & 15 \\  \hline
            Staff organizer & 4 \\ \hline
            Platform worker & 2 \\ \hline
            Warehouse worker & 1 \\
            \bottomrule
        \end{tabular}
    }
}
\caption{Our participant composition, split by workplaces and roles.}
\label{fig:composition}
\end{figure}

Our participant composition is summarized in \Cref{fig:composition}. The roles occupied by interviewees were diverse, including software engineers, infrastructure and IT workers, product managers, logistics workers, data analysts, customer support staff, QA analysts, and technical writers, as well as full-time and volunteer union staff who supported organizing campaigns at tech companies. Interviewees' workplaces included large tech companies, start-ups, non-profit organizations, media organizations, and worker co-operatives located throughout the U.S., with many based in San Francisco, New York, and Seattle. We also interviewed two worker-organizers at platform companies and three from companies sub-contracted by large tech companies.

Our inclusion criteria restricted this study to U.S.-based participants. This presents a large gap in our study as a growing portion of the labor that powers the U.S. technology industry is located in other countries~\cite{2021_catanzariti_global_labours}. Additionally, our recruitment methods may also have skewed our participant composition. Despite our expansive definitions of “tech worker” and “organizer”, our recruitment messages were still most likely to reach individuals who self-identify as both. For instance, we note that most of our participants were full-time employees at technology companies, and that we only interviewed a handful of platform or logistics workers, and no vendors (food, janitorial, transportation, etc.) who provided services to tech companies. Although we did not collect demographic data such as race or gender, 12 participants self-identified in our interviews as either a person of color, woman, and/or non-binary. We note that this is another potential gap in our study, as some participants cited workplace discrimination as a catalyzing factor in their unionization campaigns.

\subsection{Data collection and analysis}
We analyzed our data using an inductive and iterative approach guided by contextualized grounded theory~\cite{2006_charmaz_grounded}. Through rounds of coding and memoing, we surfaced high-level themes that captured tech workers’ motivations to unionize (e.g., “Layoffs/Firings,” “Better Pay and Benefits”), challenges of unionizing at tech companies (e.g., “Silos”, “Union-Busting”), unionization strategies (e.g., “Education”, “Relationship Building”), worker gains (e.g., “Community”, “Organizing Experience”), views of the tech industry (e.g., “Negative Impact on Society”, “Devaluing Tech Workers”), as well as future visions of the tech worker movement (e.g., “Building Solidarity”, “Militancy”). 

\subsection{Positionality and Ethical Considerations}
We perform this research within a particular set of social, institutional, and political circumstances that necessarily inform our work, and actively draw on our own experiences in interpretation and analysis. Three of us have been heavily involved in union campaigns in the U.S.—two of which were at technology companies—and the fourth author benefited from being a part of a bargaining unit as a member of a long-standing union. Despite our position as academic researchers, we all self-identify as tech workers. After our collective experiences on these campaigns, we were motivated to understand and improve the state of labor organizing, especially in the technology industry, but we acknowledge that these experiences undoubtedly shaped our individual perspectives on this topic. 

Due to the sensitive nature of interviews and high personal safety risks associated with labor organizing, including but not limited to the threat of workplace retaliation \cite{2022_kapoor_privacy}, we were obligated to conduct our work in a manner that does not hinder any worker's ability to safely engage in collective action. We took several measures to reduce these risks, including removing personally identifiable information such as name, residential location, and place of employment from transcripts. We also did not collect demographic data such as race or gender. All processes were approved by the appropriate Institutional Review Board. We refer to participants via ID and gender-neutral references. We also recognize the possibility that this work may reach an unintended audience that could use these findings to undermine unionization efforts. With this in mind, we chose not to present certain results which might be used for future union-busting.

\section{Findings}

In this section, we begin by exploring the motivations driving U.S.-based tech workers to unionize, including striving for better working conditions and increased input in company decision-making. Next, we examine the challenges they face in organizing an increasingly fragmented and unstable tech sector, alongside their strategies to overcome these obstacles, such as fostering community and political consciousness. Finally, we discuss their visions for the future of tech labor, emphasizing the importance of building solidarity and developing alternative organizing frameworks.

\subsection{Organizing Against Precarity in the Tech Industry}
\label{sec:motivation}

Despite the narrative that tech jobs are highly paid, flexible, and have strong benefits~\cite{2018_roberts_employment}, multiple interviewees described precarious circumstances, including discrimination, wage disparities, and mass layoffs. Participants also wanted workplace accountability and more input in company decision-making, especially when their companies engaged in unethical projects or practices. These workers organized not only to improve working conditions but also to reshape the tech industry’s culture from within.

\subsubsection{Better Working Conditions}
Amid widespread layoffs and terminations across the sector, many turned to unionization to secure stronger protections and severance terms. For P8, the firing of employees who had advocated for change within their company became a rallying point for unionization. They sought to create a \emph{“more durable organization”} to protect those who were vocal at the company. Similarly, P10 recounted that the firing of a \emph{“beloved”} coworker who spoke against management decisions left workers reeling: \emph{“We were like, `What are we going to do to protect ourselves? And what are we going to do to push back against this really bad behavior?'”} This catalyzed their union campaign with a central demand for more transparent and fair disciplinary procedures. When their company later announced mass layoffs, the union pivoted to supporting affected workers and used their bargaining power to negotiate better severance packages. Platform-based workers also used collective action to push against arbitrary suspensions without notice or due process, despite their inability to formally unionize. When P11 and others had their accounts suspended without explanation, they organized through a mutual aid group and leveraged public pressure on social media to reverse the suspensions.

Tech workers highlighted other “bread and butter” issues such as low pay, lack of benefits, and unequal working conditions as motivations. P19, a customer support worker, countered the narrative that tech workers were privileged, noting that many---such as customer support, contractors, and logistics workers---were making \$30,000 a year, which they added, \emph{“is really not a privileged position to be in.”} As a logistics worker, P49 noticed that crucial benefits such as childcare that were readily available to office workers were not extended to those in logistics. Similarly, contractors for a large tech company (P24, P35, P41) described how they did not receive the same benefits or pay as their full-time counterparts. Despite being offered a lower salary and less flexibility than at their previous job, P24 described initial excitement about working at this company: 

\begin{displayquote}
“Everything I had heard about [COMPANY] was like, ‘This is the greatest place in the world to work. There's video games and pinball machines and the food's free' [...]
And that was a total lie, basically. So I got stuck in this job that I couldn't afford to keep. I couldn't live off 19 bucks an hour as a single [parent] at that time.”---P24
\end{displayquote}

P41 further described the existence of a \emph{“two-tier system”} that positioned contractors as inferior to full-time employees. They elaborated that this system was not helpful, only creating resentment due to unequal treatment:
\begin{displayquote}

“The two-tier system, it still sort of left its mark on every aspect of the work and the work environment [...] It's hard not to feel resentful at times that you're making literally a third of what the person sitting across from you is making, and you're working together on the same project, doing the same work.”---P41
\end{displayquote}

Interviewees further pointed out that contractors, customer support, and logistics workers (who were more likely to be women, people of color, and people from other marginalized groups) were treated as fungible compared to the average engineer. Across the board, interviewees discussed a strong motivation to unionize as a result of persistent cultures of misogyny, gender-based harassment, and racial discrimination unaddressed by management, despite company narratives claiming support for diversity, equity, and inclusion. Recognizing the deep inequities within their workplace, some engineers and full-time staff organized to support their more precarious colleagues, seeing collective action as a way to advocate for fairer treatment across job classifications. Yet, as P9, a former staff union organizer, noted, building \emph{“wall-to-wall”} unions that include all workers in a single workplace remains challenging: \emph{“When their working conditions are so different, it raises this question of, what does solidarity actually mean or look like?”} These testimonials underscore the challenge of fostering solidarity across disparate working conditions.

\subsubsection{Worker Input in Company Decision-Making}
Many interviewees emphasized the need for greater input in company decision-making, citing poor management and shifting company values as key motivators for organizing. Unionization helped workers hold companies accountable to the values they publicly claimed to uphold, especially at mission-driven organizations. P25, for example, described their startup's leadership team as \emph{“very volatile”} and saw unionization as a mechanism to influence the company's direction. They added that employee resource groups (ERGs) created by management aimed to \emph{“make employees feel like they have support, but at the end of the day, didn’t increase their negotiating power.”} P27 shared a case in which their company, once supportive of abortion rights before an acquisition, revoked support for employees traveling out-of-state for essential healthcare after the overturning of Roe v. Wade\footnote{The overturning of Roe v. Wade was a landmark decision in 2022 which ended federal protections for the right to abortions, leaving abortion regulation to individual U.S. states.}. Similarly, P6, a contracted staff member at a platform company, explained how platform workers consistently faced a lack of transparency and input around platform changes that directly impacted their livelihoods. Instead, they wanted gig workers to drive company decision-making.

Although most organizing efforts---especially those led by more precarious workers---prioritized immediate material demands like fair pay and job security, many organizers saw ethical responsibility as inseparable from labor struggle. Interviewees expressed a strong desire to organize around ethical concerns related to tech's complicity in climate change, policing, and militarism, voicing resistance to the development of harmful technologies. Many of these workers cited the ongoing genocide in Gaza and drew inspiration from notable campaigns, such as Google workers protesting against Project Maven and No Tech for Apartheid ~\cite{2018_shane_maven,anonymous_2021_nimbus,2022_harrington_nota}, while also acknowledging risks of retaliation~\cite{2024_thornbecke_google}. A data worker, P28, described feeling powerless when they realized the data they labeled may have been used to train AI-based weaponry:

\begin{displayquote}
“I was watching a genocide happen on my Instagram feed, and I was horrified, because it's obvious that a lot of the tools like [Weapon System 1]  and [Weapon System 2], I helped develop, and it just infuriated me, because there's no way for me to go take my data out of that. Nobody asked me if it was okay if I helped program something that would kill people. I should be allowed to take my data back.”---P28
\end{displayquote}

Their reflections reveal not just abstract ethical concerns, but a visceral response to the material consequences of their labor. For many organizers, reckoning with their contributions to harms perpetrated by the tech industry became a motivating force for organizing, even when publicly opposing harmful technology carried personal risk. Ultimately, unionization efforts were not only driven by a desire to improve workplace conditions but also by a collective demand to hold the tech industry accountable from within, challenging its complicity in global injustice and asserting worker agency over the future of technological development.

\subsection{Organizing Challenges Across Silos}
While many of the organizing challenges described by tech workers are not unique to the tech industry, interviewees pointed to distinct factors tied to tech's culture, values, and infrastructures that made organizing difficult. These included navigating the complexities of organizing a distributed workforce, balancing transparency and security in digital labor organizing, experiencing retaliation and high turnover, and facing instability from company acquisitions and venture capital influences.

\subsubsection{Organizing a Distributed Tech Workforce}
Interviewees highlighted both challenges and opportunities of organizing remotely. Although meeting remotely allowed organizers to coordinate discreetly without being monitored by management, the absence of informal “watercooler” spaces for casual discussion created difficulties in building relationships with coworkers.
Organizers added that being remote also led to impersonal communication methods that further divided the workplace. For instance, P25 described how their company’s Slack channel became \emph{“extremely hostile”} and a \emph{“battleground,”} due to management’s ability to control its discourse, which complicated organizing efforts. P16 highlighted similarly significant barriers in fostering trust remotely, which is necessary for people to take more radical action: \emph{“People are not as willing to really go to the mat and make really radical action, strike, stand up to the boss with people who [...] they're not shoulder-to-shoulder with on a daily basis.”}

These challenges were compounded in companies with workforces distributed across multiple countries. P6, whose organizing committee spanned nearly twenty countries, described the challenges of coordinating across time zones, noting that even \emph{“a little scheduling hiccup can postpone us a whole week.”} Legal restrictions also complicated efforts to include international workers in unionization. P25 explained how organizers failed to engage their international coworkers, as U.S. labor laws prevent them from being included in bargaining units. P25 reflected that this was a \emph{“huge error,”} as it allowed management to drive a wedge between workers and to paint the union as being explicitly anti-international worker, contributing to the loss of their union campaign.

Platform workers like P11 also discussed challenges of determining shared goals with an international workforce with varying needs and expectations, especially around pay. P6, for example, suggested that while the base rate on their platform was not a viable living wage in the U.S., it could adequately support multigenerational families in other countries. P6 anticipated two concerning outcomes from the increasing precarity of tech workers: a rise in gig work as a result of mass layoffs and a growing reliance on outsourced labor from the Global South, both to exploit more workers and to suppress collective organizing: 

\begin{displayquote}
“Gigification is the biggest tool that capitalism is going to use to absolutely crush worker movements [...] There is no legal protection, there's no sort of International Labor Board or anything like that. There's not even international labor law [...] It's all about the exploitation of the Global South. That's our entire system. There is always someone you can exploit more.”---P6
\end{displayquote}

P10, a customer support worker, reflected on growing trends of outsourced labor, especially with the rise of AI. When their colleagues got laid off, P10’s company hired support workers from the Philippines to help fill gaps. Feeling that there was not enough alarm raised among their coworkers about outsourcing, they expressed equal concern with AI’s impact on labor, particularly its use to justify job cuts. AI drastically transformed the nature of P41's work as a data analyst. Complex, intellectually stimulating work was replaced by monotonous tasks as their responsibilities shifted toward data labeling for AI systems created to automate that very labor. P41 reported feeling disposable as a result of this change: \emph{“I think it's also hard to feel motivated to do work that makes yourself obsolete.”} Their disillusionment deepened upon learning that hundreds of their data labeling collaborators in India had been laid off due to AI being deemed sufficiently trained enough to replace them. As work becomes more siloed, temporary, and gigified, and workers become deskilled, our interviewees envision that the only path forward lies in building international solidarity. 

\subsubsection{Balancing Transparency and Security in Digital Labor Organizing}

While certain tools are essential for organizing, interviewees acknowledged that an over-emphasis on technical solutions and security infrastructures could hinder the ability to build strong campaigns. Several organizers discussed how tech workers tend to \emph{“over-index”} on technical tools and problem-solving rather than building relationships (P2, P22, P40). P40, an engineer, noted that although tech workers may excel at logistical tasks, they often neglect relational skills that are core to organizing: \emph{“If I'm giving advice to people building unions in tech, [it's] that the actual thing that does the work is not a skill that you'd learn as a software engineer. It's one-on-ones. It's really deeply understanding your coworkers. It’s emotions.” }

At the same time, organizers emphasized the crucial role of data in organizing, from creating promotional materials to bring awareness to workplace issues (P5), to developing disciplinary action trackers for those going through performance improvement plans (P4). A common practice in early stage union organizing is “workplace mapping,” which involves gathering data on employees’ level of support for the campaign. While P3 noted that these practices can feel intrusive to those unfamiliar with organizing, they emphasized their necessity: \emph{“Without data, you just can't effectively organize without that high level view of things.”} To secure member trust, P3 advocated for privacy-respecting approaches---such as storing data locally rather than on public platforms. However, some digital tools designed to support labor organizing raised ethical concerns. P21, for example, felt \textit{“sketched out”} by \textit{Unit}, a VC-backed platform for signing union cards. Introduced in 2020 to streamline the unionization process without needing to affiliate with a national union, the app raised questions about its for-profit model and long-term sustainability before shutting down in March 2023 \cite{2021_unit}.

Surveillance by management added another layer of complexity. Organizers recounted being acutely aware that they were being monitored by management for union activity, through corporate Slack or social media. P40 described how their company’s productivity tracking systems could easily be used to surveil union activity: \emph{“There's a database of every keystroke you've ever entered and your mouse movements and what app you were using...They could be like, `All right, who typed `union?' Who typed `organize'?' This is possible, which is scary.”} To protect against surveillance, organizers relied on non-workplace communication channels via Signal, Slack, or Discord and implemented security protocols. Organizers acknowledged the risk of creating friction, in terms of building momentum, trust, and accessibility. P3 described the challenge of balancing security with accessibility, as choosing technical tools with steeper learning curves \emph{“made it less inclusive, or made it more difficult for people who are not tech savvy to join.”} Overly burdensome security protocols also created distrust among workers. While P2 understood the need, this made them feel less trusted and risked replicating the same silos that management used to separate workers.

Others discussed tensions between security and visibility. P22 noted that \emph{“balance between security and impact”} was important, emphasizing that despite the risk of disciplinary action, building strong public campaigns was critical for building momentum. P5 similarly said that visible organizing made other workers feel more comfortable supporting the union. For example, P4 became more comfortable supporting union efforts after seeing their coworkers show support on Slack. These testimonials reveal that over-reliance on tooling and stringent security protocols in union organizing can create friction, and exacerbate existing divisions between workers. These organizers stressed the importance of prioritizing trust and relationship-building over security culture. 

\subsubsection{Retaliation and High Turnover}

Many organizers' companies retaliated against workers in response to union organizing. Several organizers observed how company leadership limited workers’ access to information as a response to union activity (P2, P5, P22). After becoming aware of union activity, P2’s company shut down their internal Slack for several weeks, indicating that leadership was \emph{“trying to break down every communication channel we have.”} 
This disconnection amid mass layoffs hindered outreach efforts as there was no way to see who was still at the company. Further, companies that had once promoted flexible remote work began enforcing stricter working conditions and return-to-office mandates. Some workers viewed these changes as a union-busting tactic and a pretext for further layoffs, as many employees were unable to suddenly restructure their lives to accommodate these changes.

While the threat of layoffs mobilized many, they also posed a significant barrier to organizing. Multiple interviewees were directly impacted by layoffs or firings at their workplaces, some perceiving it as direct retaliation for organizing. These layoffs created a \textit{“chilling effect”} on union organizing, as described by P1: \emph{“We had just done one layoff, and then another round came in fall, which made everybody scared [and] made organizing difficult, because people were worried that their name would be added to the next list.”} P31 explained that tech’s culture of individualism and meritocracy further complicates matters by framing layoffs and firings as personal failures, pushing workers to prove their worth through hard work instead of challenging the systemic causes.

The turmoil of widespread firings and layoffs weakened unionization efforts, made it difficult to build sustainable campaigns, and, in some cases, contributed to the dissolution of unions. P27 observed that not only were organizing committees targeted in layoffs, but also that trans and Black employees were disproportionately affected, pointing to systemic biases against historically marginalized groups. Workers on H-1B visas had to be cautious about publicly supporting the union out of fear of retaliation and threats of deportation. Although U.S. labor laws are meant to protect workers against retaliation for union organizing, several interviewees argued that these safeguards did little. P5, for example, highlighted that because workers are under “at-will” employment, companies can fire workers for any reason, making it difficult to prove retaliation. 
While some organizers had successfully filed unfair labor practice charges\footnote{Unfair labor practices are actions by employers or unions that violate U.S. labor laws. This includes retaliating against union activity or refusing to bargain in good faith.} for being unjustly fired and later received back pay, P2 reasoned that 
\emph{“the fee [COMPANY] got slapped on them by the Labor Board was nothing compared to the amount of money that they make every year.”}

Another tactic companies used was offering voluntary severance. After layoffs, P10 and P19’s union dissolved when each employee was offered a buyout option to “voluntarily” exit. This led the remaining organizers, as well as most employees, to leave. P19 said that many who took the seemingly lucrative buyout eventually questioned their decisions, as many remained unemployed or could not find equally well-paying jobs. Other interviewees, including engineers, disclosed that they were unemployed at the time, with some being jobless for well over a year. These experiences counter the common narrative that tech workers can job hop without needing to take a pay cut. P10 observed that this narrative previously served as an excuse by workers who did not want to organize. P10 challenged this view, noting that as the systemic issues pervasive in the industry are inescapable, it is better to invest in making their current environment better: 
\begin{displayquote}
“People are very much like, ‘Well, I don't want to get involved because I'm just already looking for something else’ [...] I always have this thing where I want to scream, ‘But where are you going to go?’ It's not better anywhere.' [...] The only thing you can do is build a better community where you're at.”---P10

\end{displayquote}

These testimonials illustrate that, despite enduring corporate retaliation and navigating a climate of fear that undermined collective efforts, organizers rejected the notion that they could walk away from poor workplace conditions. Instead, they recognize that these issues are systemic to the industry and require a sustained commitment to changing them from within. 

\subsubsection{Instability from Company Acquisitions and Venture Capital Influences}
Many interviewees highlighted the obstacles of organizing smaller companies under threat of acquisition or backed by venture capital (P1, P18, P25, P33, P34). P25 observed that venture capital actively suppressed union activity, due to investment agreements that required company leadership to attest that there was no union activity within the organization. After their union campaign went public, the company’s funding was withdrawn, resulting in a drastic decrease in company valuation and low employee morale, as many were financially invested in the company.
According to P25, company leadership leveraged this loss to crack down on union support: \emph{“We lost a ton of people straight away and management hammered that home: 'We lost this funding, we don't have much runway. This company is going to shut down. If you vote for union, you probably are going to kill the company.'”}

Another tactic used to undermine unions is restructuring through company sales. Several interviewees (P9, P17, P19, P27) provided examples where unions were effectively dissolved after acquisition. Often, personnel were sold as assets rather than stock, allowing the acquiring company to treat workers as new hires and not have to recognize the union. P27 detailed how the sale of their company allowed the new ownership to lay off most of their workforce, effectively dissolving their union and leaving workers without severance pay or worker protections.

The structural challenges posed by venture capital and acquisitions point to broader limitations in traditional labor organizing methods in the tech industry. P44 criticized the \emph{“shop-by-shop”} model to union organizing, arguing that it fails to keep up with the tech industry's volatile ownership structures. In an industry that prioritizes rapid growth and short-term returns over sustainable business practices, organizers expressed a desire to adopt alternative organizing strategies to build a stronger movement.

\subsection{Worker Gains and Losses}
\label{sec:gains}
Across the board, interviewees described the immense emotional toll of union organizing, especially in the face of stressful workplace conditions, tensions in workplace relationships, layoffs, firings, and union losses. P16 compared their unionization journey and subsequent layoff to a \emph{“heartbreak.”} Others expressed feelings of burnout and even trauma. Reflecting on these challenges, P25 admitted that they \emph{“would pause before doing it again”}:

\begin{displayquote}
“Do I have the energy? Am I willing to lose this job? Am I able to perform at the same level while doing this extra work? What are the odds of winning? I would want to say that I would do it again if given the opportunity. But you know, it's not an easy decision.”---P25
\end{displayquote}

This highlights the immense personal sacrifices that tech worker-organizers often make, risking their livelihoods and emotional well-being in pursuit of better working conditions and workplace accountability. Despite these challenges, many interviewees highlighted the concrete wins they achieved through their efforts, including increased salaries, 
better healthcare, job security protections, expanded PTO, wellness stipends, workplace accommodations, as well as stronger protections against discrimination and intrusive monitoring, among others. However, it was the intangible gains, such as community building, organizing experience, and political consciousness that interviewees felt were equally, if not more, significant in shaping their identity as tech workers and organizers. 

\subsubsection{Breaking Down the Silos and Building Community}
The deep sense of community formed through organizing was consistently brought up by interviewees as one of the most rewarding aspects of unionizing, especially in contrast to their siloed workplaces. Many workers described how organizing connected them with people within the company whom they would not otherwise have engaged with. Despite being remote, P6’s organizing committee cultivated strong bonds, deep trust, and mutual support that extended beyond collaboration on union activities. P6 stated they felt supported in ways they have never experienced in other workplaces: \emph{“There's no substitute for that feeling of complete trust. I trust these people. I've not even met two of my [co-organizers] in person, but I trust them so deeply [...] I feel like they see me and I see them.” 
} Similarly, P16 reflected on how the connections made through unionization efforts led to \emph{“transformative”} personal growth and building a sense of shared struggle and mutual support, despite coming from diverse backgrounds: \emph{“We tried and failed to do a lot of things, but I feel like people were able, and willing, in so many ways on their journey, [to stand] up for one another.”}

Building community also extended beyond the workplace. Multiple interviewees named the Tech Workers Coalition (TWC) and the Labor Notes Conference as two important avenues to connect with workers inside and outside the tech industry and across the globe. P52 described TWC as an important community space where tech workers gather, find resources, and support one another, noting that this space was intentionally created as a result of tech work being highly atomized and isolating. The Labor Notes Conference serves a similar role as a broad grassroots network of labor organizers across different sectors and unions. P33 highlighted the value of meeting members of a nursing union at the conference who had secured algorithmic transparency in their contract, noting how it underscored the real-world consequences of tech work. Additionally, P41 said that their experience at Labor Notes was life-changing because they were able to meet people who did not fit the “tech bro” trope, including women, people of color, and disabled individuals. They believed that this diverse representation was a more accurate depiction of the tech industry: \emph{“It's a disservice to the industry to depict it as this sort of white, cis-het male monolith, because that's not what the industry looks like...I think unions are some of the best ways for those marginalized voices to be amplified.”} These spaces not only helped P41 realize that they were not alone in their struggles, but also gave them hope that, through the simple act of sharing space, listening, and talking to each other, tech workers could break through silos and collectively build a better future: \emph{“There's a community there, even if we can't always see it.”}

\subsubsection{Gaining Organizing Skills}
Even after leaving a job, many organizers brought the skills they learned into new workplaces. P3, for example, got involved in their university’s union campaign and felt that the organizing skills they learned through tech were directly transferable, elaborating that \emph{“experience begets more experience.”} P1 added that they felt \emph{“galvanized”} from their experience, recognizing how far they have come since their initial engagement with unionization. As someone who had previously been a supporter of political causes rather than an active organizer, P1 described how their involvement marked the first time that they enacted direct material change over their daily realities:

\begin{displayquote}
“That gave me fuel, that differences can be made, even if they're small, even if it's within a small organization. And if we're being told that the world works one way, there are alternatives that can make it work differently. And to see that happen, and especially to watch the process through almost start-to-finish was really valuable.”---P1
\end{displayquote}

Participating in unionization was often the first organizing experience for many tech workers. P17 described it as \emph{“helpful baby organizing”} because it made room for small, tangible wins that helped them realize both their individual and collective power, which can be used for future organizing with broader social impact. P31 described training organizers as the \textit{“most important thing in the union drive, arguably more important than winning an election is training people to be excellent organizers and being able to move people to action.”} Together, these experiences demonstrate that the knowledge learned through union efforts extended far beyond the original workplace. Wherever they ended up, workers applied these skills to new contexts to continue advocating for better working conditions and catalyzing others, creating ripple effects beyond the tech industry. 

\subsubsection{Building Political Consciousness}

One of the most important takeaways from interviewees’ experiences organizing in the tech industry, was breaking free from its narrative of prestige. P1 argued that it’s \emph{“easy to think you're at the top,”} but that this sense quickly faded amid mass layoffs and stagnant wages as companies report record profits: \emph{“That high-tech salary might be great, but it might still be a sliver at the end of the day of what you collectively have worked towards. And maybe we all deserve more than a sliver.”} P16 hoped that the mass layoffs and treatment of workers as replaceable assets throughout the tech industry will “\emph{help provide a sense of consciousness of our mutual precarity.}” Others began to challenge the artificial boundaries between different roles that enforced the siloing of workers. For example, P41 challenged the binary distinction between white- and blue-collar workers, asserting that they have more shared experiences and struggles than differences: \emph{“White-collar and blue-collar aren't mutually exclusive, binary groups that one never moves from...We all have a lot more in common than we think.”}

This growing political consciousness inspired interviewees (P5, P6) to read more about the broader labor movement's political history, helping them understand that their struggles are part of a longer legacy of workers fighting for justice. Through this perspective, P5 recognized that the systems companies use to exploit workers are deliberately constructed, and therefore can be dismantled: 
\emph{“If you know what's happening and you know how it's being manufactured to happen in a particular way, then it can also be destroyed...I think that did save my life. And I think that can save other people's lives.”} P6 echoed the need to learn from labor history to help connect tech workers’ current struggles with other sectors: \emph{“Tech work is not divorced from manufacturing, is not divorced from any other career pathway.”} P6 referenced early Ford factory workers and their mass injuries, siloing, deskilling, and replaceability due to the assembly line as a cautionary tale around the misuse of AI to fix social problems. P40 directly challenged techno-optimism, claiming that it has not delivered on its promises to improve people’s lives due to its prioritization of corporate greed, while also giving no control to workers:

\begin{displayquote}
“We had this…utopian vision, maybe 20 years ago, of like, tech is going to save the world [...] Instead we've gotten, like, gilded poo [...] Every single person I've worked with, with maybe a few exceptions, is motivated to help users and to build good products and to do good things. And so why is it that everyone doing the work wants these things, but the outcomes are bad? And the answer is capitalism, the answer is management, the answer is greed.”---P40

\end{displayquote}

In response to this disillusionment, some workers transformed their relationship with their work. Learning about the inherent disposability of tech workers, P2 realized that they should not give their whole selves to their workplace: \emph{“I'm more than my work…at any time, employers could get rid of me. So why am I giving 200\% to my employer, when they think of me at a 50\% value add to the company?”} For P5, organizing gave them a deeper sense of purpose than their job ever could: 

\begin{displayquote}
    “I am a tech worker, and I produce things for my coworkers and for my stakeholders. And I'm like, this does not matter. This means nothing, I could disappear, this job could disappear [...] And then on the flip side, I get to do all this organizing work where I get to produce literally a pamphlet, like one single sheet of paper. And that can change the trajectory of many people's lives.”---P5

\end{displayquote}

P10 likened this growth in political consciousness as a \emph{“light being turned on inside a person.”} As workers connected their struggles to a broader movement and legacy, they brought this political awareness into new contexts. \emph{“The light went on inside so many people, and they're taking that with them wherever they go.”}

\subsection{Future Visions}
\label{sec:visions}
Despite the challenges of unionizing in the tech industry, interviewees remained steadfast and hopeful about the future of the tech worker movement. They noted a significant shift in the conversations surrounding tech worker organizing, and growing interest in unionization. As the tech worker movement becomes more mature and established, interviewees stressed the importance of shifting towards building a long-term future.

P19 argued that the concept of a future is inherently linked to unions, stating that \emph{“without a future, a union cannot exist.}” They elaborated that unions are intentionally designed to bring long-term sustainability into a capitalistic industry that prioritizes short-term profits. P19 contended that capitalism actively undermines futures by promoting narratives of dystopia and hopelessness, which discourage workers from investing in long-term goals. P19 therefore viewed unions as a means to reclaim better futures, asserting, \emph{“I truly believe that unions and organizing help you create a future while they're trying to take our future away from us.” }

To build a stronger, more sustainable labor movement, organizers expressed the urgency of building solidarity with workers across sectors and internationally. Noting the limitations of existing labor-organizing frameworks, they imagined new possibilities for tech unions. This often involved the need to look back in order to look forward. 

\subsubsection{Building Solidarity}

Recognizing the limitations of U.S.-based labor laws to protect workers against the volatility of the tech industry, interviewees highlighted the necessity of moving beyond the traditional \emph{“shop-by-shop”} model of union organizing. 
Organizers referenced examples of sectoral-based collective actions that could serve as models. For example, P2 and P18 drew inspiration from the Writers Guild and SAG-AFTRA strikes, where workers in a broad sector with significant pay disparities came together to organize ~\cite{2023_koblin_writer_strike, 2025_sagaftra}. Similarly, P14 envisioned a future where tech work can be supported through trade union structures, complete with union halls, apprenticeships, and worker-managed job pipelines, offering a more stable alternative to the volatility of capital-driven employment.

At the local level, P41 recognized the inherent power of being affiliated with a local branch of a national union based in their city, which represents over 1,000 cultural and education workers, as well as tech workers: \emph{“The cultural impact that we can have on the city would be kind of immeasurable.”} Likewise, P1 called attention to service workers striking with teachers at Pratt in 2019~\cite{2019_pratt_seiu} as an example: \emph{“More union power is better than less, even if it’s not your union, even if it’s not your grievances. I’d love to see more tech spaces, say, ‘Here's where UAW’s next picket is, let's join it.’”} P5 echoed this call, advocating for tech workers to build solidarity, not just across the tech supply chain, but also across sectors: \emph{“If there was a walkout from the tech workers on top of a walkout from the warehouse workers, all of [COMPANY] would shut down...And so you can just weave this large web of how things can build off of each other.”} As a logistics worker, P49 appreciated seeing the show of support from office-based tech workers for their unionization campaign and welcomed opportunities for further collaboration. While they did not self-identify as a “tech worker,” they acknowledged the superficial divisions between logistics workers and engineers, alongside a growing mutual understanding of their shared challenges and demands.

In addition to building worker solidarity, interviewees emphasized the importance of going beyond workplace issues to connect with local communities and broader social movements. P49 highlighted that collaborating with immigrant community-based organizations was crucial for building their campaign, due to many logistics workers being immigrants themselves, and that too often the community aspect is not taken as seriously as it should be within the labor movement. Further, P44 emphasized the importance of drawing strategies from broader social movements, which have been less focused on establishing material victories and more about \emph{“shifting the weather”} to become more favorable for concrete victories in the future. Reflecting on the 2018 Google walkouts, in which 20,000 employees worldwide participated, P44 observed that although the walkouts did not secure immediate material gains, they were instrumental in inspiring tech workers across various companies to unionize. They further argued that effective organizing should function as an interconnected ecosystem rather than as isolated efforts, with looser activist-and social movement-based organizing working in tandem with more structured organizing to push for material change~\cite{engler_this_2016}. 

Interviewees also acknowledged the importance of building transnational solidarity. P6 envisioned multinational collective actions in solidarity with vulnerable workers in other countries:\emph{ “If we had a strike of all the [COMPANY] workers in California, and all the workers in China, that would make a big difference.”} While P6 acknowledged that it's currently illegal under U.S. labor law for workers at the same company to strike in multiple countries, they also said, \emph{“that doesn't mean we can't do that,”} reflecting their belief that labor laws should not be seen as a barrier to collective action, especially when demanded by working conditions. These testimonials demonstrate a growing urge among tech worker-organizers to build solidarity across divisions. As tech workers increasingly recognize their interdependence with workers in other sectors and countries, there is a growing awareness of the limitations of traditional labor frameworks. 

\subsubsection{Alternative Frameworks}
Some organizers saw the need to push the boundaries of unions and envision alternative frameworks that accommodate a wider range of tactics and worker experiences. Several interviewees expressed frustration with the limited impact of existing tactics, especially those that require unfavorable compromises. As a result, some tech workers revealed a growing urgency for more militant organizing within the tech industry. P3, for example, argued that\textit{ “striking is the only way to get results,”} adding that workers do not necessarily need a union to strike, but that stakes are higher without one. P39 echoed this, calling attention that \textit{“militant unionism,”} despite its higher risks, offers the potential for greater wins. They further recommended pushing back against the status quo of a traditional union structure. P40 too stressed the need for experimentation in navigating an unstable and unpredictable landscape~\cite{cait2025playbook}: 

\begin{displayquote}
    “We have to try stuff, and we're in an era of tech labor organizing where we don't know what the tactics that will work are [...] that part of the work is to experiment and figure out what the hell is going on and what is effective. We don't know in the same way that people of a century ago were figuring out what tactics work, and we got to be talking to each other.”---P40
\end{displayquote}

Platform workers like P28 have had to explore alternative forms of organizing, such as mutual aid groups, given their exclusion from traditional union structures. Despite these constraints, P28 emphasized that they still possess meaningful collective power to resist unethical uses of technology and demand greater control over their labor: \emph{“We still have collective power…. A lot of people will say collective bargaining. I don't want to bargain. I'm not talking about bargaining. We have power. We are the users, and we are the creators, and we have the power.”} According to P47, a staff union organizer, minority unions in Big Tech such as Alphabet Workers Union remain valuable even without collective bargaining power, as they bring \textit{“education and experience of class struggle into the heart of Big Tech,”} where forming a majority union has proven extremely difficult. Recognizing these challenges, P52 became a volunteer organizer with the Emergency Workplace Organizing Committee (EWOC), which aims to train and support non-unionized workers through grassroots organizing.

Several interviewees (P5, P6, P39, P40, P43, P53) noted that today’s tech worker-organizers can draw valuable lessons from labor history to navigate the rapidly changing landscape of the tech industry. P6 explained how Luddites destroyed mechanized factory equipment during the Industrial Revolution to resist the intensification of worker exploitation caused by these technologies, describing a possible inspiration for modern-day resistance against AI. P6 also cited the Black Panther Party as another source of inspiration for integrating labor activism into everyday life, with its active presence in churches, schools, social clubs, and support groups. P43, a QA analyst, highlighted that historically, the most marginalized workers---such as garment workers and immigrant farm laborers---have led labor organizing because they had the most at stake. Today, P43 observed that QA testers---the lowest-paid workers in the gaming industry---are playing a similar leading role in tech unionization efforts~\cite{yahr2025wisconsin}.

Others (P6, P22, P27) expressed hope for a future where the tech industry can fully embrace worker-cooperative models, where profits are shared and decision-making is democratized. This vision is not just a distant aspiration, as represented by P45, a unionized worker-owner of a tech cooperative whose position involves balancing the interests of both workers and owners equally. Although this model poses challenges in terms of scalability and funding \cite{2014_cheny_coops}, it has allowed workers like P45 to promote accountability within their company, prioritize worker wellbeing, and uphold shared values, offering a concrete example of how a more equitable tech industry might be realized.


\section{Discussion}
Our findings foreground the experiences of U.S.-based worker-organizers who have advocated for better working conditions in an increasingly stratified and precarious tech industry. While some campaigns won improved pay, protections, and visibility, others were thwarted by siloed workplaces and retaliation. Despite these setbacks, many tech workers described organizing as a source of profound transformation, enabling them to build community and foster political consciousness. Situating their struggles within broader historical contexts, they began to challenge the dominant logics of the tech industry and envision a stronger, more resilient tech worker movement rooted in cross-sectoral and international solidarity. As CSCW researchers, we can draw valuable insights from these efforts. In this section, we revisit three interrelated themes: how the tech industry disempowers worker organizing (5.1), how organizers build solidarity across boundaries (5.2), and how CSCW can align itself with the tech worker movement (5.3). 

\subsection{How the Tech Industry Disempowers Workers}

In our interviews, tech workers described the many persistent challenges that make collective organizing in the tech industry difficult. These barriers are not only structural, but also ideological, shaped by forces that influence corporate practices as well as organizers' tools, strategies, and beliefs. Across the industry, tech jobs are increasingly stratified and precarious as tech firms adopt platform-based business models that rely heavily on subcontracting and outsourcing to cut labor costs and avoid employment responsibilities \cite{2019_rahman_thelen_platform_business}. The high volatility of venture capital-backed firms undergirds precarious work arrangements, as companies are incentivized by investor-driven pressures for rapid growth and profitability to employ contingent labor \cite{2024_shestakofsky_startup}. Meanwhile, AI-driven restructuring and task automation have further contributed to mass layoffs and the deskilling and devaluation of workers, affecting even highly paid engineers \cite{narayan_ai_work_2024, so2025cruel}.

Workers in more precarious positions---such as contractors, gig workers, and those on H-1B visas---described facing heightened risks when organizing due to unstable employment and limited protections. These material differences often translated to divergent organizing goals, with some workers prioritizing basic protections and others seeking values-based participation in organizational governance. Restrictive labor laws often hinder the unionization of contractors, gig workers, and other contingent employees who make up a significant portion of the workforce. Finally, fragmented legal frameworks across different states and countries pose further challenges, making it difficult to adopt universal strategies for organizing.

While these dynamics are materially grounded, our interviews also reveal the enduring influences of ideological forces, such as those associated with the \textit{Californian Ideology}, in shaping how workers and employers make sense of and respond to these transformations. We find traces of these ideological logics in both employer strategies and worker beliefs alike, highlighting values of entrepreneurialism, meritocracy, and innovation while obscuring class stratification, labor exploitation, and opposition to organizing efforts~\cite{barbrook1996californian, tarnoff2020techworker}. For example, workplace tools like Slack were often used by management for surveillance and discipline---actions not unique to the tech industry, but nonetheless resonant with the ideology's preference for a kind of individualized accountability that frames unions as antagonistic and unnecessary. Employment Resource Groups (ERGs) allow workplaces to project a progressive image without enacting meaningful change. Similarly, stock options and buyout packages were deployed to incentivize higher-paid workers to opt out of organizing, appealing to the belief that advancement stems from individual talent rather than collective struggle. This mindset also shaped the affective attachments workers formed to their jobs, leading them to perceive layoffs and firings not as systemic failures but as personal shortcomings, which discouraged organizing efforts~\cite{so2025cruel}. 

We also observed how some organizers, especially those in technical roles, gravitated toward solutions that emphasized optimization, data tracking, or security protocols, sometimes at the expense of trust-building. These tendencies reflect an internalization of workplace norms that echo the CI’s valorization of technical mastery and rational design over collective deliberation. Believing that data-driven workplace practices could be easily translated and applied in organizing contexts, some organizers inadvertently created silos, eroded trust, and prioritized technical expertise over relational organizing. ~\cite{2020_khovanskaya_datatools,2023_khovanskaya_cyberunion}.

In short, while structural forces such as atomization and platformization contribute to the precarity of tech workers, ideological forces associated with the Californian Ideology reinforce these structures of exploitation by casting them as natural or inevitable. In this way, CI serves not as a root cause but as an interpretive frame---a set of widely shared narratives that can make structural inequality harder to name, contest, or imagine otherwise. Yet, our interviews also point to moments of ideological rupture. Workers increasingly questioned the promises of upward mobility, tech exceptionalism, and managerial benevolence, particularly as rolling layoffs, pervasive surveillance, and wage stagnation undermine these assumptions.

For CSCW researchers, this analysis carries important implications. Even in technologies ostensibly designed for workplace communication or coordination, deeply embedded assumptions about workers, power, and value shape their functionality and effects. As we turn our attention to designing systems for organizing and worker empowerment, we must ask: which cultural logics do our systems reproduce? Which values do they assume? Are those values aligned with the labor futures we hope to support?

\subsubsection{Workplace Communication Platforms as Sites of Surveillance}

Prior CSCW scholarship has examined how algorithmic management and surveillance suppress worker autonomy \cite{lee2015working, zhang2022algorithmic, anwar2021watched, kellogg2020algorithms, sannon_2022}. Building on this literature, our findings surface how seemingly ordinary enterprise workplace communication platforms such as Slack can be abused by employers to surveil workers, violate workers' privacy and their protected right to organize in the U.S.~\cite{2022-nlrb-surviellance}. At the same time, they also highlight how workers tactically respond to and appropriate these same platforms to support their organizing efforts. Recall how P4 became more comfortable supporting union efforts after seeing many of their peers visibly show support for the union on their workplace Slack. Other examples show how workers repurposed Slack's emoji feature to subversively respond to their employer's read-only messages ~\cite{2022-mizon-nyt} and developed platform counter-strategies against union-busting tactics ~\cite{union-busting-tcgplayer}. These cases point to the tactical possibilities within mainstream workplace communication technologies when appropriated by organizers for counter-power, even as these platforms are used by employers for surveillance and control. We argue for greater attention and engagement by the CSCW community to the potential conflicting uses of such platforms and their tensions. This includes exploring opportunities to support worker privacy and collective organizing in the design and governance structures of these platforms, including supporting end-to-end encryption for private work channels \cite{aclu_encryption_2023}, more flexible access management to limit managerial oversight on certain channels and data \cite{wired_teamsprivacy_2023}, and clear and transparent data and surveillance policies \cite{2022_kapoor_privacy}.

\subsubsection{Technosolutionism in Labor Organizing}
Our findings in 4.2.2 also point to how worker-organizers can unintentionally reproduce technosolutionist logics when they “over-index” on digital tools, data practices, and security protocols. Consistent with prior HCI and CSCW research, we find that ICTs can undermine organizing efforts—reinforcing hierarchies, introducing accessibility barriers, or shifting focus toward efficiency and rigid data practices rather than relationship building \cite{2019_ghoshal_computing_grassroots_tech, ghoshal2020toward}. These systems are often repurposed enterprise software designed for managerial efficiency---logics often at odds with the values of grassroots movements. Yet they remain prevalent due to their accessibility, affordability, and ease of maintenance, particularly when alternatives are resource-intensive or require advanced technical capacity \cite{chan2019promise, lampinen2018member}. 

Drawing from Ghoshal’s~\cite{2019_ghoshal_computing_grassroots_tech} design tenets of inclusivity, privacy/security, and social translucence, we suggest pathways for reorienting organizing technologies to better support relational labor practices. This involves adopting tools with lower barriers of entry, prioritizing accessibility for the most technologically marginalized members, and critically evaluating whether chosen tools align with the values of the labor movement. While privacy and security are important in organizing contexts, we echo P22’s need to find a \emph{“balance between security and impact.”} As seen in the \#TechWontBuildIt campaign—where workers mobilized on platforms like Twitter to oppose militarism—and in the rise of Game Workers Unite, which used Discord and Facebook to organize across geographic boundaries, digital technologies have played a crucial role in building momentum and visibility for the tech worker movement \cite{selander2016digital}.

\subsection{Organizing Across Borders and Boundaries}

Labor organizing in the tech industry is facing an unprecedented moment of change and challenge. The question posed by P9,\emph{“When their working conditions are so different…what does solidarity actually mean or look like?”}—lies at the heart of understanding broad-based labor organizing in this sector. As the industry continues to evolve rapidly, traditional union strategies in the U.S., particularly the \emph{“shop-by-shop”} model, are increasingly inadequate in addressing the challenges faced by tech workers. In response, our interviewees expressed a desire for new forms of organizing—ones that draw on grassroots networks and broader movements that can adapt to diverse labor conditions and transcend the institutional and legal boundaries of traditional organizing.

\subsubsection{Supporting Transnational Labor Organizing}

While interviewees strongly desired to coordinate more militant collective actions— such as cross-sectoral or international strikes—these efforts are hindered by both structural and legal barriers. Nonetheless, such actions are occurring within the tech industry. One notable example is the 2018 Google Walkouts, where employees from over 50 cities worldwide organized a mass protest in just a few days~\cite{chiu2018google,wong2019google}. While prior work has emphasized the importance of informal networks in localized labor organizing, particularly for gig workers who cannot formally unionize \cite{thuppilikkat2024union, qadri2021mutual, qadri2021s, abilio2021struggles}, our findings suggest that these structures are equally critical at the international level. As Niebler ~\cite{niebler2023transcending} notes, transnational solidarity within the tech industry largely takes shape through informal and grassroots forms of organizing, lacking institutionalized structures.

The Tech Workers Coalition (TWC) emerged repeatedly in our interviews as a key space for this type of informal organizing. TWC is a transnational network with global and local chapters, connecting workers across geographies while responding to localized labor issues. Workers reflected on how its low barrier to entry, focus on mutual support, and emphasis on community events and political education make it particularly effective at cultivating organizing capacity among workers who might otherwise be excluded from traditional union pathways. TWC~\cite{2024-miller-twc} and other labor networks~\cite{2023-nota-twc} have been critical in supporting transnational and even formal labor campaigns, such as Labor for Palestine~\cite{labor4palestine} and SEIU’s effort to unionize janitorial workers~\cite{seiu-janitors}. 

As CSCW researchers, there is an opportunity to not only design tools to support decentralized and grassroots forms of organizing, as many others have done \cite{irani_2013, salehi2015salehi, harmon2019rating}, but also to complement and bridge efforts together towards more structural interventions that can drive policy change at the international level. For example, Fairwork~\cite{fairwork-about} is a transnational action-research project and online platform co-developed by researchers from 38 countries alongside workers, platforms, unions, policymakers, and activists that aim to support platform workers. The project evaluates platforms and employers based on principles, such as fair pay, fair conditions, and fair management. To date, this initiative has pushed over 60 companies to improve their worker policies, and even gained the attention of the International Labour Organization (ILO), which has demonstrated interest in establishing international guidelines informed by these principles.

\subsubsection{Drawing Strategies from Broader Social and Labor Movements}

Rather than existing in isolation, the tech worker movement also draws strength and strategic insight from historical and contemporary labor struggles. The success of the Writers Guild and SAG-AFTRA strikes showed interviewees that it is possible to mobilize workers across disparate pay scales, job roles, and levels of precarity. Learning about the Luddites and the Black Panther Party provided our interviewees with alternative models for militant action and community-based forms of resistance. These references suggest that tech workers actively see themselves not as isolated cases, but as part of a longer lineage of workers and activists fighting for justice. This recognition aligns with calls for more historically grounded approaches to CSCW that might inform the way we design technologies \cite{soden2023historicism}. Notably, Sabie et al.~\cite{2023-sabie-luddites} also draw inspiration from the Luddites on unmaking as a form of technological emancipation. Similar arguments for refusal and destruction have increasingly been taken up across the field, as scholars seek to confront the harms posed by dominant technologies such as biometric surveillance in the workplace, emotion AI, among others~\cite{2023-roemmich-emotionai, awumey2024systematic}. We can look to the Luddites for inspiration for modern-day resistance tactics against these systems. This may include dismantling and abolishing algorithmic management systems used for surveillance and control \cite{sum2025itslosinggameworkers}, designing AI technologies to be worker-centered \cite{2020_fox_worker, zhang2023stakeholder}, and developing tools to support collectively reporting or tracking workplace harms \cite{irani_2013, calacci_pentland_2022}. 

\subsection{The Role of CSCW in the Tech Worker Movement} 
In “The Making of the Tech Worker Movement,” Ben Tarnoff ~\cite{tarnoff2020techworker} argues that the contentious and contradictory “middle layers” that tech workers are situated in as part of the “virtual class” can be ripe ground for organizing. We extend this insight to CSCW researchers who also occupy these “middle layers,” not as external observers but as entangled participants within the tech ecosystem. As tech researchers, we often move between academic, industry, and even activist spaces, collaborating across them while navigating their contradictions. This positionality grants us a rare opportunity and responsibility to build stronger solidarities with tech workers and materially support their organizing efforts. In this section, we unpack how CSCW researchers can support the tech worker movement within and beyond academia. Second, we reckon with our own labor power as a CSCW community in shaping more just futures for tech labor.

\subsubsection{Supporting the Tech Worker Movement Within and Beyond Academia}

Just as the organizers in our study had to confront internalized contradictions between their roles as tech workers and as organizers, CSCW researchers must also contend with similar tensions. While CSCW and HCI have increasingly engaged with questions of tech labor, this research is often shaped by Global North politics and institutions, and entangled with some of the same ideological and structural forces associated with the tech industry \cite{raval2022considerations, irani2016stories}. As a result, even well-intentioned scholarship can inadvertently reproduce the very dynamics it seeks to critique: emphasizing technological solutions over political struggle \cite{2022_ahmed_future}, centering managerial or corporate perspectives \cite{2019_fox_managerial, 2020_khovanskaya_datatools}, or treating precarious workers—especially those in the Global South—as peripheral subjects or illustrative cases rather than agents of knowledge and resistance \cite{raval2022considerations, 2021-raval-invisibility, irani2016stories}. These practices risk disconnecting research from the lived realities of those most impacted by technological harm, and may inadvertently expose them to further precarity, surveillance, or retaliation. To address this, we must critically examine how we conduct research and the political and epistemic frameworks that shape it \cite{2023_tang_cscw_labor, clarke2020research}: \textit{Who do we center in our work? Who ultimately benefits from it? What structural forces shape how we study labor? Rather than solely focus our gaze on those most vulnerable, how might we direct our analytical gaze upward to
make structures of power visible, while holding companies and institutions accountable for systemic harms?}

As scholars and organizers have emphasized, the tech worker movement must be led by the most precarious workers---warehouse workers, gig workers, and others whose labor is foundational to the industry and who have led its most militant actions \cite{niebler2023transcending,tarnoff2020techworker,2019-what-we-learned-nedzh-tan}. CSCW researchers must follow their lead, not only by documenting their struggles but by aligning our research agendas, partnerships, and institutional resources with their demands and visions for change. In line with the political traditions of participatory design \cite{bannon2018reimagining}, researchers and workers have been turning to digital workers' inquiry that reorients design practice toward collective empowerment \cite{gallagher2025digital, calacci_2022, holten2024uniting, nyman2024towards, selander2024data} The Data Workers Inquiry project is just one example, where data workers from five different countries served as community researchers to surface their labor conditions, while academic partners provided support in data collection and analysis. Such projects demonstrate how CSCW research can redistribute resources and amplify worker-led knowledge production.

Additionally, our interviewees emphasized the importance of organizer training, skill-building, and the development of political consciousness alongside a desire for stronger transnational ties. As a global and interdisciplinary community, CSCW researchers are well-positioned to contribute to this work \cite{irani2016stories}. Political education initiatives might include co-creating zines on organizing strategies \cite{collectiveaction2022layoffguide, collectiveaction2024pip}, collaborating with workers to host workshops on building solidarity and collective action in computing \cite{raval2022considerations, huber2022solidarity, 2023_tang_cscw_labor}, co-developing digital security training for organizers \cite{2022_kapoor_privacy}, or designing curricula that connect ethical questions in technology to historical and contemporary labor struggles---both local and transnational \cite{widder2023pwer, twc2022teachin}.

\subsubsection{Recognizing Our Own Labor Power as CSCW Researchers}

We also invite the CSCW community to consider its own collective power. In examining activism within the AI community, Belfield \cite{belfield2020activism} notes that this loosely defined group holds power both as an epistemic community and as a site of labor organizing. This dual positioning—researchers both as knowledge producers and workers—has enabled AI practitioners to form coalitions, challenge unethical practices, and push for institutional change. The CSCW community holds similar potential.

Taking inspiration from Computer People for Peace in the 1970s, a modern coalition of CSCW researchers could similarly form a bridge between academic inquiry and political action to oppose militarism, racism, corporate control in computing, and demand accountability from industry and academia while advancing more just technological futures \cite{huber2022solidarity, irani2019patron}. Signs of this are already emerging. Unionization efforts across higher education in the U.S. are growing, led by graduate students, adjunct faculty, and postdoctoral researchers~\cite{2024-salvatory-unionization-higher-ed}. These efforts reflect a growing recognition that academic labor, like tech labor, is shaped by precarity, power asymmetries, and extractive logics. This work is particularly urgent amid the rise of tech oligarchy and escalating attacks on U.S. institutions of higher education~\cite{2025_highered}. For CSCW researchers, acting in solidarity means supporting organizing efforts within our own institutions, resisting exploitative research practices, and forming broader coalitions. 

Looking ahead, the future of tech labor and CSCW’s role in it requires a fundamental reimagining of work, technology, and justice. As P19 observed, \emph{“Without a future, a union cannot exist,”} highlighting the vital role unions play in enabling workers to collectively envision and build more just futures, while resisting the imposition of pre-determined trajectories that erode hope and long-term sustainability. Dominant “future of work” narratives often position AI and automation as inevitable, reinforcing the techno-utopian determinism of the Californian Ideology and masking the labor exploitation and power asymmetries that undergird it \cite{2022_ahmed_future}. While CSCW has offered critical interventions into the present and near-future harms of these trajectories, much of this work remains reactive and focused on mitigating the consequences of systems already in motion \cite{to2023flourishing, kim2024envisioning}. In contrast, our interviewees show that the labor movement is not only a site of resistance, but also a force for articulating and enacting affirmative visions of futures, grounded in collective care, equity, and sustainability. One example is the \emph{Worker as Futurist} project, a paid writing workshop in which Amazon workers crafted speculative fiction envisioning “The World After Amazon”~\cite{haiven2024}. Reclaiming the future from corporate techno-utopianism, the project illustrates how organized labor can advance worker-led approaches to technological design and policy. 

Recognizing labor organizing as an essential act of “futuring”~\cite{2023_vertesi_futuring} demands a broader, more speculative research agenda. Drawing on Greenbaum's \cite{1996_greenbaum_labor} ``Back to Labor,'' and building on recent efforts that center worker wellbeing and their collective power \cite{spektor2023designing, irani_2013, silberman2018responsible, calacci_pentland_2022, huber2022solidarity, 2023_tang_cscw_labor, 2022_ahmed_future}, we argue for a shift away from designing individualized responses to technological harm. Instead, we propose moving from a focus on \textit{Future of Work} to the \textit{Future of Labor}: a future in which CSCW engages with labor not as a passive object of technological change, but as an active, world-making force capable of shaping sociotechnical futures through struggle, solidarity, and structural transformation.

\section{Conclusion}
We currently stand amid a resurgence of collective action in the technology industry. Facing increasing precarity and lacking levers for ethical accountability, tech workers across workplaces, industries, and international boundaries are fighting for control over the conditions of their work and the outputs of their labor. Through 44 semi-structured interviews with U.S.-based tech workers who have contributed to this movement, we come to understand a cross-section of tech worker-organizers' values, motivations, and challenges, as well as their visions for the future of tech organizing. As academic researchers of computing technologies, we call on the CSCW community to stand in solidarity with these workers, organizing together to collectively create a more ethical and accountable technology industry. 

\section{Acknowledgements}
We are deeply grateful to the 44 tech worker-organizers who generously shared their experiences with us. We are indebted to the Tech Workers Coalition and Collective Action in Tech for their support and for the important work they do in strengthening the tech labor movement. We also thank Tucker Rae-Grant, JS Tan, Linda Huber, David Widder, Shivani Kapania, Franchesca Spektor, Jordan Taylor, Christine Mendoza, and anonymous peer reviewers for their thoughtful feedback.

\bibliographystyle{ACM-Reference-Format}
\bibliography{bibliography}

\end{document}